# High Current Density in Monolayer MoS$_2$ Doped by AlO$_x$

Connor J. McClellan[1], Eilam Yalon[1,2], Kirby K.H. Smithe[1], Saurabh V. Suryavanshi[1], and Eric Pop[1,3,*]

[1]*Electrical Engineering, Stanford University, Stanford, CA 94305, USA.*
[2]*Present address: Electrical Engineering, Technion, Israel Institute of Technology, Haifa, 32000, Israel*
[3]*Materials Science and Engineering, Stanford University, Stanford, CA 94305, USA.*

**ABSTRACT:** Semiconductors require stable doping for applications in transistors, optoelectronics, and thermoelectrics. However, this has been challenging for two-dimensional (2D) materials, where existing approaches are either incompatible with conventional semiconductor processing or introduce time-dependent, hysteretic behavior. Here we show that low temperature (< 200°C) sub-stoichiometric AlO$_x$ provides a stable *n*-doping layer for monolayer MoS$_2$, compatible with circuit integration. This approach achieves carrier densities >2×10$^{13}$ cm$^{-2}$, sheet resistance as low as ~7 kΩ/□, and good contact resistance ~480 Ω·μm in transistors from monolayer MoS$_2$ grown by chemical vapor deposition. We also reach record current density of nearly 700 μA/μm (>110 MA/cm$^2$) in this three-atom-thick semiconductor while preserving transistor on/off current ratio > 10$^6$. The maximum current is ultimately limited by self-heating and could exceed 1 mA/μm with better device heat sinking. With their 0.1 nA/μm off-current, such doped MoS$_2$ devices approach several low-power transistor metrics required by the international technology roadmap.

The success of modern electronics has relied on conventional silicon transistor scaling, enabling advancements in computing technology year after year for over five decades. Recent field-effect transistors (FETs) have used top-down fabrication to realize ultrathin silicon "fins" (*i.e.* FinFETs)[1,2] for improved control of leakage current and performance. However, these approaches have limitations imposed by process variation and the degradation of silicon mobility in ultrathin, approximately sub-4 nm layers.[3] The emergence of sub-1 nm thin monolayer 2D semiconductors could therefore extend transistor scaling, representing the ultimate limit of semiconductors, without an analogue among bulk materials like silicon.[4] For example, monolayer 2D semiconductors like MoS$_2$ could enable sub-5 nm scale transistors[5] and, owing to their direct band gap, could also allow integration with optoelectronic devices.[6] Recent research has demonstrated the integration properties of monolayer 2D semiconductors, including three-dimensional (3D) monolithic systems[7] and flexible electronics.[8]

However, the atomic thinness of 2D semiconductors has raised questions about the ability to dope them and, consequently, about their ultimate performance in integrated circuits. While doping bulk materials like silicon is achieved with substitutional impurities, such an approach in a three-atom-thick material could significantly degrade the mobility. In addition, performance cannot only be judged based on (low-field) mobility, but also on the maximum drive current $I_{on}$, because circuit delays are proportional to $CV/I_{on}$, where $C$ is the capacitance including parasitics and $V$ is the voltage. To meet International Roadmap of Devices and Systems (IRDS)[9] specifications for low-power transistors, the on-state current must exceed $I_{on} > 480$ µA/µm, while the off-state current must remain $I_{off} < 0.1$ nA/µm, ideally within a narrow voltage swing (e.g. 0.7 or 1 V) for low-power operation.

Past studies of doping 2D materials have explored surface functionalization of acceptor[10,11] or donor[12] states, but a large degradation in transistor subthreshold swing (*SS*) was often observed due to induced trap states. Chemical charge transfer doping was also proposed,[13,14] however such approaches face stability and integration issues. Instead, sub-stochiometric metal oxides (like $MoO_x$ for *p*-type[15] and $AlO_x$ or $TiO_x$ for *n*-type[16,17]) have been used as *stable* doping layers of 2D materials, but doped devices typically experience a severe reduction in on/off current ratio ($I_{on}/I_{off}$) and poor *SS*. Such challenges have ultimately prevented the achievement of 2D transistors with high on-current and good $I_{on}/I_{off}$. Furthermore, most doping studies on 2D semiconductors have been limited to thicker, multi-layer semiconductors,[13,18] which have no clear technological benefit over ultrathin silicon-on-insulator (SOI) or FinFETs,[1,19] far more mature technologies.

In this work, we demonstrate a stable doping approach that preserves the transistor $I_{on}/I_{off}$, while enabling record-high $I_{on}$, record-low sheet resistance, and low contact resistance in a three-atom-thick semiconductor. These results are enabled by the increase in carrier concentration from doping while maintaining a low interface trap density through annealing. Importantly, this is achieved with $MoS_2$ grown by large-area chemical vapor deposition (CVD),[20] which is necessary for practical applications. Our three-atom-thick $MoS_2$ transistors reach $I_{on} \approx 700$ µA/µm at 5 V (~300 µA/µm at 1 V) while maintaining $I_{off} < 0.1$ nA/µm. These achievements advance monolayer semiconductors to an important position for low-power logic and memory, approaching industrial specifications.

**RESULTS AND DISCUSSION**

**Device Design:** Figure 1a shows the schematic of our transistors fabricated using monolayer $MoS_2$ grown by CVD directly onto $SiO_2$ ($t_{ox} = 30$ nm) on $p^{++}$ Si, which also serves as a back-gate, with pure Au contacts[21] (also see Methods). The uncapped (and undoped) monolayer $MoS_2$ has a field-effect mobility of 35 to 40 cm$^2$V$^{-1}$s$^{-1}$ in this work, which could range from 30 to 50 cm$^2$V$^{-1}$s$^{-1}$ in our previous studies on similar CVD-grown material.[20] To dope these, we first use electron beam evaporation to deposit a thin 1 nm Al seed layer that immediately oxidizes upon air exposure to form sub-stoichiometric AlO$_x$, followed by 15 nm of AlO$_x$ deposited by atomic layer deposition (ALD), and additional details are given in Methods. Figure 1b displays an atomic force microscopy (AFM) image of multiple such MoS$_2$ devices in a transfer length method (TLM) structure with channel lengths from $L$ = 180 to 980 nm, as measured. For good contact resistance ($R_C$) estimates, such TLM structures must include channel lengths ranging from "short" (dominated by their contacts) to "long" (dominated by the channel resistance).[21] Extrapolating $R_C$ only from long channel devices could lead to large uncertainty and even apparently negative contact resistance from TLM extractions.[22] The corresponding photoluminescence (PL) and Raman spectra of these channels before and after AlO$_x$-capping are displayed in Figs. 1c and 1d, respectively, and additional details are provided in Supporting Section S1.

**Doping *vs.* Trapping Induced by the Oxide:** Before presenting the electrical data, we note that doping the 2D material by metal oxides can result from at least two distinct processes. In the first process the charge is induced by trap states at the semiconductor/oxide interface (classically referred to as $D_{it}$) or in the oxide near the interface (*e.g.* border traps).[23,24] These traps are energetically located within the energy gap of the 2D semiconductor and ultimately lead to degradation of mobility or *SS*. The other process for doping the 2D material is by transfer of electrons or holes from states that do not overlap with the energy gap of the 2D semiconductor, analogous to modulation doping in high electron mobility transistors (HEMTs).[25] Similar effects have been attributed to dipoles in high-*k* dielectrics on Si transistors, where dipoles affect the mobile carrier density in the channel.[26] In this case, the induced charge carriers end up in the conduction (valence) band for *n*-type (*p*-type) doping, and do not degrade the *SS* or mobility of the 2D transistor. Such techniques are commonly used in the semiconductor industry to adjust the threshold voltage ($V_T$) in Si transistors, utilizing either fixed charge or dipoles.[27]

Figures 2a,b show measured linear and logarithmic drain current *vs.* gate voltage of a 3 μm long MoS$_2$ channel before AlO$_x$ deposition (gray), immediately after AlO$_x$ doping (light red), and after an anneal in N$_2$ at 200°C for 40 minutes (dark red). Note that all *I-V* measurements shown here include forward and backward sweeps, as labeled by small arrows, while the minimum and maximum gate voltages are limited by the breakdown field of the gate dielectric. Immediately after ALD of AlO$_x$ the carrier and current density increase, but the *SS* and transconductance ($g_m = \partial I_D/\partial V_{GS}$) degrade, indicating the as-deposited AlO$_x$ leads to carrier trapping. The induced trap density is high, $\Delta D_{it} \sim 5\times10^{13}$ cm$^{-2}$eV$^{-1}$ estimated from the change in *SS* (Supporting Section S2). However, after annealing in N$_2$ the *SS*, mobility, and $g_m$ recover to their values measured in the undoped channel, with a negative $V_T$ shift corresponding to ~$8.6 \times 10^{12}$ cm$^{-2}$ electron doping (Supporting Section S5), and a current increase by >50% at the highest $V_{GS}$ shown. The sheet resistance of this long channel after doping and annealing is the lowest reported to date, $R_{sh} \approx 7$ kΩ/□, estimated after subtracting the small (<4%) contribution of the contact resistance discussed below.

The trapping and doping states observed in Figs. 2a,b are linked to AlO$_x$ defects and their energy distribution at or near the AlO$_x$/MoS$_2$ interface.[28-30] Figure 2c displays an energy band diagram of the doping effect, showing three defect states in AlO$_x$ modeled previously using density functional theory (DFT),[30] originating from oxygen vacancies in sub-stochiometric AlO$_x$. These defect states each have a charge and energy level that depend on the electron occupation. Defects with high electron occupation have lower energy levels and no charge (D$^0$), whereas removing electrons raises the defect energy level and leaves positive charge. The change in defect energy level from removing or adding electrons has been shown with DFT to occur from redistribution of the surrounding atoms in the metal oxide, changing the required energy to add or remove an electron, respectively.[28-30]

Shallow defects that have energy levels within the MoS$_2$ band gap and available electron states (D$^{1+}$) lead to trapping of MoS$_2$ channel electrons, decreasing their mobility because the localized electrons "hop" between defects.[31] If the defects donate their electrons and reside at energy levels above the MoS$_2$ conduction band (D$^{2+}$), the MoS$_2$ electrons are not trapped. These higher energy defect states donate electrons and become positively charged, inducing negative (mobile) charge in the MoS$_2$ channel. Remote Coulomb scattering with these charged D$^{2+}$ states could limit the channel mobility, as is the case for HEMTs,[32] but this is not observed here as the MoS$_2$ mobility after AlO$_x$ capping and anneal is virtually unchanged from the uncapped samples. This indicates that such remote Coulomb scattering is either screened by the AlO$_x$ or that the MoS$_2$ mobility is more strongly limited by intrinsic defects and phonons in our samples.[33]

Using the Stanford 2D Semiconductor (S2DS) FET model,[34] we successfully simulate the data in Fig. 2b. The model can describe both the sub-threshold (diffusion) and above-threshold (drift) current components, with additional details in Supporting Section S3. The large degradation of *SS* immediately after AlO$_x$ deposition is due to mid-gap defects, included in the model as an interface capacitance ($C_{it} = q^2 D_{it}$)[27] which reduces the overall gate capacitance ($1/C_G = 1/C_{ox} + 1/C_{it}$) where $C_{ox} = \epsilon_{ox}/t_{ox}$, $\epsilon_{ox}$ and $t_{ox}$ being the permittivity and thickness of the SiO$_2$, respectively. Thus $C_G < C_{ox}$, which can lead to an overestimation of carrier density and underestimation of mobility (Supporting Section S3).

The good agreement between the experimental data and the $D_{it}$ model shows how trapping and doping can be induced by changing the energy level and defects density in the AlO$_x$. Annealing in a non-reactive, inert N$_2$ ambient[35,36] after AlO$_x$ deposition helps promote the defects to donate electrons to the MoS$_2$, analogous to dopant activation steps in conventional semiconductors. These results highlight the difference between trapping and doping of mobile charge in 2D materials. Here we achieve 2D doping without degradation of $I_{on}/I_{off}$, unlike previous studies where the decreased $I_{on}/I_{off}$ was (incorrectly) attributed to large doping, although this was likely a $D_{it}$ effect.

We also fabricated long channel ($L$ = 6 μm) top-gated MoS$_2$ FETs using the AlO$_x$ as the top gate insulator (Supporting Section S4). As expected, we found that after anneal (the doping state), the AlO$_x$ enables good electrostatic control and low gate leakage, evidence of an insulating oxide. However, before anneal (in the AlO$_x$ trapping state) the top gate control was weak and the gate leakage current was much higher (see Supporting Section S4). The leaky AlO$_x$ in the trapping state is consistent with defect states within the MoS$_2$ band gap, which lead to trap-assisted tunneling and gate leakage.

We have analyzed other sub-stoichiometric oxides for doping 2D materials in previous studies, including MoO$_3$ (for $p$-doping WSe$_2$ and graphene),[15,37] as well as TiO$_x$ and NiO$_x$.[38] However, we have found that AlO$_x$ provides the best results for $n$-type doping likely due to the Al seed not reacting with nor damaging MoS$_2$. In contrast, Ti and Ni can react with and damage monolayer MoS$_2$,[38] respectively, leading to lower mobility.

**Contact and Sheet Resistance:** To obtain the contact resistance of the doped MoS$_2$, we use TLM structures[21] from Fig. 1b and measure resistance *vs.* length (Fig. S5). Figure 3a displays the effective mobility ($\mu_{eff}$, here an average over the six channels) from the sheet resistance, before and after the N$_2$ anneal. The mobility increases from 12.8 cm$^2$V$^{-1}$s$^{-1}$ before the anneal (due to the large $D_{it}$) to 33.5 cm$^2$V$^{-1}$s$^{-1}$ after the anneal, similar to that of our undoped monolayer MoS$_2$.[20] The average sheet resistance is $R_{sh}$ = 9.0 ± 0.5 kΩ/□ at $n \approx 2\times10^{13}$ cm$^{-2}$ in this TLM and ~7 kΩ/□ in the doped long-channel device of Fig. 2a. These are the lowest sheet resistances observed to date for monolayer MoS$_2$ at room temperature, comparable to those achieved using superionic conductor (LaF$_3$) gating at the lower temperature of 220 K.[39] Figure 3b shows the contact resistance *vs. n*, reaching as low as $R_C \approx$ 480 Ω·μm for Au with monolayer MoS$_2$ after AlO$_x$ doping. This is also the lowest contact resistance to any CVD-grown monolayer semiconductor and one of the lowest amongst all 2D semiconductors.[16,40]

We attribute the low $R_C$ to the reduction in Schottky barrier width between Au and MoS$_2$ with increased carrier concentration, although the AlO$_x$ doping layer only touches the edge of the contact.

This reduction in $R_C$ from channel doping has been observed before[15,16] and can be attributed to two causes. First, as the $R_{sh}$ of the metal is far less than that of the 2D material, the current transfer length ≈ 50 nm (*i.e.* region of current injection under the contact, see Supporting Section S5), leading to most of the current being injected very close to the contact edge.[41,42] Second, due to the 2D nature of the channel, the depletion region of the Schottky contact extends beyond the contact edge into the channel.[41] As a result, increasing the carrier density by $AlO_x$ doping at or near the edge of the metal contact reduces the Schottky depletion region, increasing tunneling from metal to semiconductor as is observed in highly doped Si/metal contacts. This is further evidenced by comparing this result to our previous $R_C$ results in $MoS_2$ devices,[21,43] as shown in Figure 3b, where the new reduction in $R_C$ is achieved by reaching higher carrier densities through the use of $AlO_x$ doping.

The low $R_C$ and $R_{sh}$ in our monolayer $MoS_2$ allow us to reach a maximum current $I_{on}$ ≈ 690 μA/μm in a 380 nm long channel at $V_{DS}$ = 5 V (Fig. 3c), achieving a record current density $J_{on}$ > 110 MA/cm$^2$ for the three-atom-thick $MoS_2$ with $t_{ch}$ = 6.15 Å.[44] This current density is the highest recorded to date in a 2D semiconductor, approximately 5× higher than the typical electromigration current densities of common metals and surpassed only by that of graphene (a 2D semimetal)[45] and carbon nanotubes[46] which are near ~1 GA/cm$^2$. Strictly speaking, this device reaches $I_D$ ≈ 300 μA/μm at $V_{DS}$ = 1 V and the IRDS low-power specification[9] (> 480 μA/μm) is met at $V_{DS}$ = 2 V. To meet this at ≤ 1 V would require a two-fold improvement in mobility or a combined increase in mobility and carrier density, coupled with a reduction of contact resistance. We also keep in mind that the device shown in Fig. 3c has $L$ = 380 nm, whereas the IRDS specifications[9] are meant for FETs of 10-20 nm channel length.

Figure 3d shows the transfer characteristics of the same transistor, demonstrating stable doping with only slight degradation after 60 days in air, and negligible hysteresis (both forward and backward measurement sweeps are shown). The device can turn off to 0.1 nA/μm and exhibits $I_{on}/I_{off}$ ≈ 2.5 × 10$^6$ despite the high doping even at $V_{DS}$ = 5 V (Supporting Section S6), contrasting other doped 2D material transistors where the high on-state current was only achieved with low on/off ratio. These metrics are comparable to or better than those of recent silicon-on-insulator (SOI) devices,[47] yet achieved in a ~16 times thinner monolayer $MoS_2$ channel, and additional comparisons are provided in the Benchmarking section below. We emphasize that reaching high on-current density while preserving *SS* and sufficient $I_{on}/I_{off}$ is a critical figure of merit to benchmark practical doping techniques of 2D materials.

**Current Density Limits:** Despite the record current density achieved here in a three-atom-thick semiconductor, it is important to ask what is limiting the maximum current and whether this could be improved further, given that high current (per transistor width) is required for high speed circuit operation. With a fixed parasitic resistance of 960 Ω·μm (= $2R_C$) and channel resistance of 300 Ω·μm, our devices could reach 600 μA/μm at 20 nm gate lengths and $V_{DS}$ = 0.75 V, meeting IRDS low-power specifications.[9] In comparison, state of the art Si or III-V transistors can reach >1 mA/μm,[1,48] but in much "thicker" channels.

The maximum current density of a transistor is limited by mobility or saturation velocity, carrier density, contact resistance, and self-heating during operation. Naturally, higher mobility (such as in III-V semiconductors or graphene) automatically leads to higher current density, but high carrier density can compensate for a lower-mobility semiconductor (as in this work). Shorter channel transistors can also reach higher operating currents, down to channel lengths that are limited by their contact resistance and injection velocity.[49] However, with a "given" set of material and contact parameters, we find that self-heating ultimately limits the maximum current achieved in our devices.

To understand this, we turn to Fig. 4, which compares our measurements (symbols) with simulations (lines) including velocity saturation,[43] contact resistance, and self-heating effects (further details provided in Supporting Section S7). By including these effects, our simulations capture the deviation from linearity in Fig. 4A, at high drain bias $V_{DS}$ > 2 V. These devices heat up significantly during measurement due to the high current density and relatively high thermal resistance of the $MoS_2$-$SiO_2$ interface and the $SiO_2$ substrate.[50] Consequently, as the input power ($P = I_D V_{DS}$) increases, the temperature rise $\Delta T$ degrades the carrier mobility and saturation velocity. We estimate a $\Delta T$ ~ 400 K channel temperature rise at the highest bias point probed here, with heat flow being limited by the relatively low thermal boundary conductance between $MoS_2$ and $SiO_2$ (TBC ~ 15 MWm$^{-2}$K$^{-1}$).[50]

We also compare the measurements with our simulations in Fig. 4b. The simulations including self-heating (solid lines) faithfully reproduce the experimental data, while simulations without self-heating (dashed lines) reach much higher current. The measured transconductance ($g_m = \partial I_D/\partial V_{GS}$) decreases with increasing gate voltage $V_{GS}$. In transistors based on typical bulk semiconductors (*e.g.* Si or III-Vs) such behavior is attributed to either contact resistance or mobility degradation from increased surface scattering at the higher transverse electric fields.[27,51] However, $2R_C$ only accounts for ~26% of the total resistance even at $n$ = 2×10$^{13}$ cm$^{-2}$ for this $MoS_2$ device, and we find no degradation of mobility with increasing $V_{GS}$ (Fig. 3a), as the electrons are already highly confined within the three-atom-thick semiconductor. The $g_m$ decrease is also more pronounced at higher $V_{DS}$, *i.e.* higher input power, indicating that self-heating effects limit our $MoS_2$ transistor performance at high fields.

Simulations without self-heating (dashed lines in Fig. 4b) reveal this transistor could reach $I_{on} \approx$ 1.2 mA/μm at $n \approx 2 \times 10^{13}$ cm$^{-2}$ and $V_{DS}$ = 5 V. These findings are consistent with recent studies of velocity saturation in monolayer MoS$_2$,[43,52] underlining that self-heating dominates the measured high-field behavior, and suggesting that other reports of high current in 2D transistors are also limited by self-heating.[16,18,53] (Also see Supporting Section S7.) We note that even with thinner insulating substrates (here the SiO$_2$ is only 30 nm), thermal dissipation is limited by poor heat transfer across the weak van der Waals MoS$_2$-SiO$_2$ interface, which is equivalent in thermal resistance to ~90 nm of SiO$_2$.[50] Thus, future efforts must consider improving heat dissipation in 2D transistors or operating them in a transient regime that is faster than typical thermal time constants,[52] which are sub-nanosecond for the 2D material[54] and of the order 10-100 ns for the transistor including its gate and dielectrics.[55]

**Benchmarking:** Figure 5a compares our results with other reports of high current in 2D semiconductors (from monolayer to ~14 nm thick) and various doping studies, including a fully-depleted Si nanowire FET.[56] Here we normalize the current by the conduction area [$J_{on} = I_{on}/(Wt_{ch})$], as opposed to only channel width to account for the channel thickness $t_{ch}$, which is a key limiter in transistor scaling.[57] Figure 5a reveals that while many 2D transistors display good on/off ratio, they lack the current density to compete with high-performance Si technology. By achieving high current drive in atomically thin, monolayer MoS$_2$ we surpass the current density of Si nanowires while maintaining good electrostatic control, highlighting the large current density and excellent electrostatics of 2D semiconductors.

These are the monolayer semiconductor transistors with the best current density reported to date, approaching IRDS low-power requirements[9] both in terms of $I_{on}$ and $I_{off}$. We also compare our results with other 2D material doping studies by plotting $I_{on}$ vs. $I_{on}/I_{off}$ in Fig. 5b. As before, we emphasize that doping methods of 2D materials should not only be evaluated on the basis of $R_{sh}$, $R_C$, and $I_{on}$, but also on $SS$ and $I_{on}/I_{off}$, as the inability to turn off 2D FETs could be an indication of charge traps. Figure 5b shows that most doping methods can induce high current in 2D semiconductors, but often by sacrificing $I_{on}/I_{off}$. As with our observations of trapping vs. doping using AlO$_x$, the doping methods that display low $I_{on}/I_{off}$ could be introducing a substantial number of mid-gap traps. Furthermore, many of these methods have not been applied to 2D monolayers (as in this work), which are electrostatically more favorable and represent the "ultimate atomic limit" of semiconductors.

Finally, in Fig. 5c we compare the $I_{on}$ achieved in this work and previous studies of monolayer MoS$_2$ as a function of transistor channel length, at the same $V_{DS}$ = 1 V and maximum $V_{GS}$ reported. Solid curves represent a simple model with $I_D = V_{DS}/(LR_{sh} + 2R_C)$ where $R_{sh} = (qn\mu)^{-1} \approx 8.1$ kΩ/□ is the average channel sheet resistance with $n = 2.2\times10^{13}$ cm$^{-2}$, $\mu = 35$ cm$^2$V$^{-1}$s$^{-1}$, and $R_C = 1$ kΩ·μm (achieved previously,[5,43] gray line) or 480 Ω·μm (achieved in this work, black line). The horizontal red lines show the IRDS low-power (LP) and high-performance (HP) requirements (at a more aggressive 0.75 V in 10-20 nm gate length FETs).[9] It is evident that micron-scale devices are limited by their mobility, but short channels ($L < 2R_C/R_{sh}$, especially < 100 nm) are strongly limited by their contacts. Thus, we expect that the largest improvements of short-channel MoS$_2$ transistors will be achieved by further reducing the contact resistance, together with reduction of equivalent oxide thickness (EOT) which will allow lowering $V_{GS}$. Other benchmarking data on multi-layer and other 2D semiconductor transistors are summarized on a new website[58] recently launched while preparing this manuscript.

**CONCLUSIONS**

We have demonstrated the doping effect of sub-stoichiometric AlO$_x$ on monolayer CVD-grown MoS$_2$. By activating dopants and reducing trap densities, we achieved record transistor current of nearly 700 μA/μm at 5 V (~300 μA/μm at 1 V), limited primarily by self-heating due to large current densities. The doping achieved with AlO$_x$ is stable, also yielding excellent sheet resistance (down to 7 kΩ/□) and contact resistance (down to 480 Ω·μm) for monolayer MoS$_2$ without degrading mobility or subthreshold swing. In contrast, previous 2D material doping methods often induced large density of interface traps that limit on- and off-state current. These interface traps can also lead to an overestimation of carrier density and underestimation of mobility. Future work should focus on selective doping near contacts, doping of the channel for threshold voltage control, *p*-type doping to enable CMOS and reducing the gate oxide thickness for lower gate voltages of high-performance transistors and 2D circuits.

**METHODS**

**MoS$_2$ FET Fabrication**: Monolayer MoS$_2$ was deposited using a chemical vapor deposition (CVD)[20] process directly onto $t_{ox} = 30$ nm of thermal dry SiO$_2$ on p$^{++}$ Si substrate (electrical resistivity of 1 to 5 mΩ·cm), which acts as a global back-gate. The MoS$_2$ was first etched into ~2 μm wide rectangular channels using electron beam lithography (EBL) and a XeF$_2$ etch. Source and drain electrical contacts were defined using EBL with channel lengths varying between 180 nm and 3 μm. Pure Au contacts of 35 nm thickness were deposited on the MoS$_2$ using electron beam evaporation at high-vacuum (~8×10$^{-8}$ Torr) followed by lift-off in acetone and isopropyl alcohol cleaning.[21] We stress the importance of using pure Au contacts to MoS$_2$ for a clean contact interface, compared to metals that oxidize or react with the monolayer MoS$_2$. Fabricating Au contacts without Ti or Cr adhesion layers requires careful processing, applying very little agitation to the sample during lift-off, and utilizing the pure Au only for the contacts and leads, not the large probing pads (which do have a ~3 nm Ti adhesion layer). The large probing pads (away from the device channels) were a stack of 20 nm SiO$_2$, 3 nm Ti and 40 nm Au, with the additional SiO$_2$ to limit leakage current from the 200 × 200 μm pad area to the substrate. If these steps are carefully followed, our liftoff yield is about 70% for the Au contacts *vs.* the Ti/Au probing pads.

**Measurements:** All electrical measurements in this work were performed in the dark and under vacuum (<10$^{-5}$ Torr) using a Keithley 4200-SCS parameter analyzer, in a Janis ST-100 probe station, at room temperature. We scratched through the AlO$_x$ layer on top of the electrical pads using the W probe tip to make electrical contact with the Au. All plotted *I-V* data shows both forward and backward sweeps, indicating minimal hysteresis in our devices. For measuring the transistor on/off current ratio, we divide the maximum on current ($I_{on}$) by the minimum off current ($I_{off}$) over the whole gate voltage sweep. Raman and PL data were taken using a Horiba Labram with 532 nm excitation laser.

**AlO$_x$ Deposition**: For the AlO$_x$ capping and doping layer, an Al seed layer was first deposited on the MoS$_2$ devices by electron beam evaporation at a base pressure of ~4×10$^{-7}$ Torr. After exposure to air, the Al seed layer immediately oxidizes into AlO$_x$. Supporting Section S8 shows the Al seed layer doping effect on MoS$_2$ FETs by applying several cycles of 1.5 nm Al deposition and oxidizing in air for 2 hours. We use a 1 nm Al layer to seed ~15 nm of AlO$_x$ deposited by ALD at 150°C with a trimethylaluminum precursor and H$_2$O oxidizing step. The Al seed layer promotes the nucleation of ALD AlO$_x$ for complete coverage of the MoS$_2$.[59] We also found that hydrogen annealing can effectively reduce the AlO$_x$, increasing the trap density and doping (Supporting Section S9).

## ASSOCIATED CONTENT

**Supporting Information:** Available free of charge at https://pubs.acs.org/doi/xx.xx/acsnano.xxx.

Raman data of MoS$_2$ capped by AlO$_x$ without Al seed layer; modeling subthreshold swing (SS), charge trapping and doping; top-gate transistor measurements; transfer length method (TLM) measurements for contact and sheet resistance; transistor model with and without self-heating; effect of Al seed layer thickness on AlO$_x$ doping; effect of H$_2$ anneals.

**AUTHOR INFORMATION**

**Corresponding Author:** *E-mail epop@stanford.edu

**Author Contributions:** C.J.M., E.Y. and E.P. conceived the experiments and wrote the manuscript with input from all authors. K.K.H.S. grew the MoS$_2$. C.J.M., S.S., and E.P. developed the subthreshold and drift current models. C.J.M. fabricated the devices. C.J.M. and E.Y. performed electrical, PL, and Raman characterizations.

**Competing interests:** The authors declare that they have no competing interests.

**ACKNOWLEDGMENTS**

Fabrication and measurements were performed in part at the Stanford Nanofabrication Facility (SNF) and the Stanford Nano Shared Facilities (SNSF), which received funding from the National Science Foundation (NSF) as part of National Nanotechnology Coordinated Infrastructure Award ECCS-1542152. This work was supported in part by ASCENT, one of six centers in JUMP, a Semiconductor Research Corporation (SRC) program sponsored by DARPA, and the Stanford SystemX Alliance. C.J.M. acknowledged support from the NSF Graduate Research Fellowship. K.K.H.S. acknowledged partial support from the Stanford Graduate Fellowship (SGF) program and NSF Graduate Research Fellowship under grant no. DGE-114747.

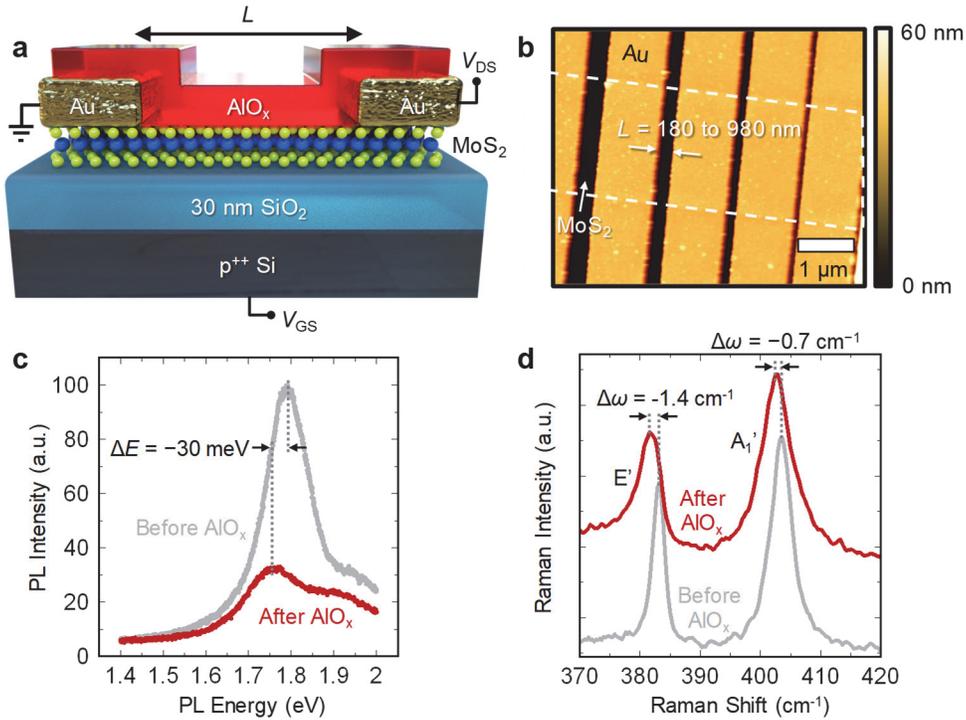

**Fig. 1. AlO$_x$ doped MoS$_2$ FET. (a)** Schematic of FET with Au contacts, 16 nm AlO$_x$ capping, and monolayer MoS$_2$ channel grown on $t_{ox}$ = 30 nm SiO$_2$ with a p$^{++}$ Si substrate as a back-gate. **(b)** Atomic force microscopy (AFM) image of transfer length method (TLM) structures used for extracting contact and sheet resistances. **(c)** Photoluminescence (PL) measurements of MoS$_2$ before and after AlO$_x$ deposition and N$_2$ annealing, showing a decrease of intensity and slight red-shift in PL peak position after AlO$_x$ deposition. **(d)** Raman spectra of MoS$_2$ before and after AlO$_x$ deposition. The AlO$_x$ deposition induces a red-shift and asymmetry of the E' mode, consistent with the Fano effect of high doping,[60,61] while the red-shift and peak broadening of the A$_1$' mode has also been correlated with MoS$_2$ doping.[62] The corresponding full-width half-maximums (FWHW) before AlO$_x$ are 3.2 cm$^{-1}$ (6.1 cm$^{-1}$) for E' (A$_1$') which then increase after AlO$_x$ deposition to 6.8 cm$^{-1}$ (7.3 cm$^{-1}$) for E' (A$_1$').

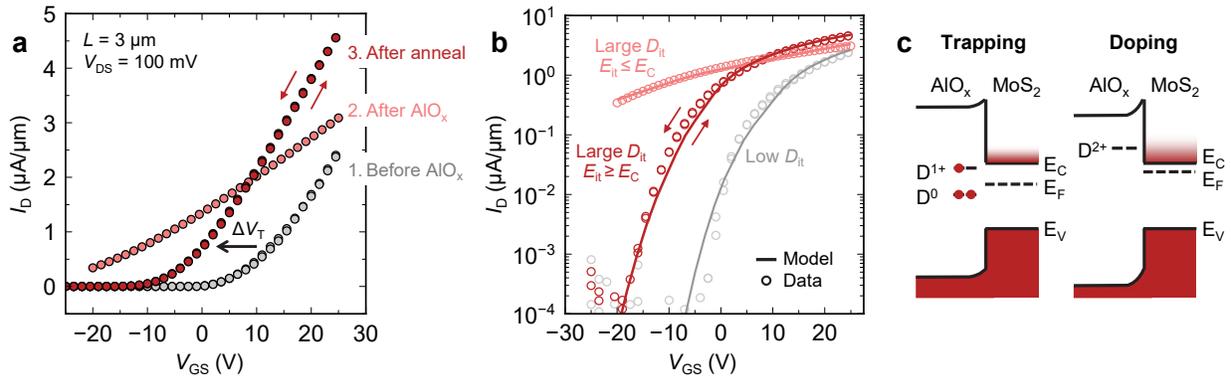

**Fig. 2. Trapping and doping in MoS$_2$ FET. (a)** Linear and **(b)** log scale measured transfer characteristics of MoS$_2$ FET before AlO$_x$ deposition (gray), after AlO$_x$ deposition (light red) and after 200°C N$_2$ anneal for 40 minutes (dark red). The -12 V shift in $V_T$ after N$_2$ annealing indicates an induced negative charge density of ~8.6×10$^{12}$ cm$^{-2}$ in the MoS$_2$. Interface trap model fitting shown in (b) matches the experimental data, demonstrating how mid-gap trap states can pin the MoS$_2$ Fermi level and reduce gate control. Small arrows show both backward and forward measurements (dual sweep), revealing minimal hysteresis. **(c)** Schematic band diagram of the trapping and doping states of the AlO$_x$/MoS$_2$ interface, D$^0$ being a defect with no charge, D$^{1+}$ a defect with 1 positive charge, and D$^{2+}$ a defect with 2 positive charges. Band bending in the AlO$_x$ is shown schematically to illustrate presence of charge.

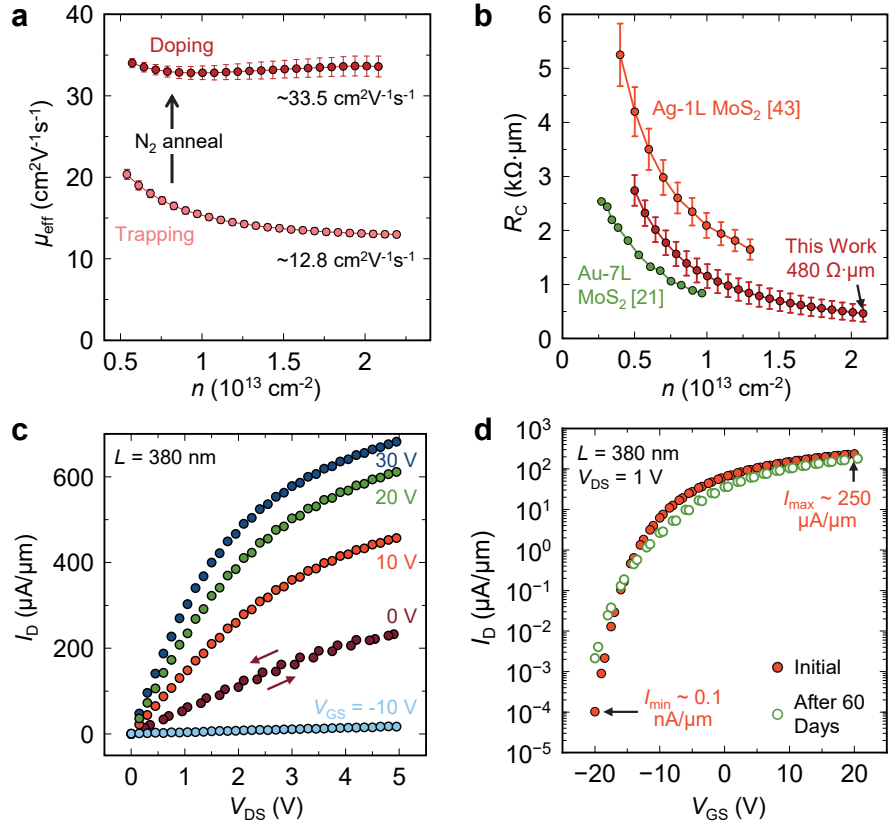

**Fig. 3. Electrical characteristics in trapping and doping states of MoS$_2$ FET. (a)** Effective mobility ($\mu_{eff}$) and **(b)** contact resistance ($R_C$) extractions *vs.* electron concentration with the TLM. The doping state (dark red) demonstrates higher $\mu_{eff}$ than the trapping state (light red), as the reduction in traps yields a lower sheet resistance of the MoS$_2$. By reaching higher carrier density, our highly doped Au-1L MoS$_2$ FETs demonstrates lower $R_C$ than previously measured $R_C$ between Ag-1L MoS$_2$ and Au-7L MoS$_2$.[21,43] **(c)** Measured output and **(d)** log-scale transfer characteristics of a 380 nm long AlO$_x$ doped MoS$_2$ device, reaching nearly ~700 µA/µm while maintaining a high on/off ratio of ~10$^6$. The doping method is stable with only slight degradation after 60 days in air, as shown in (d). Every *I-V* shows forward and backward measurements (small arrows), with minimal hysteresis.

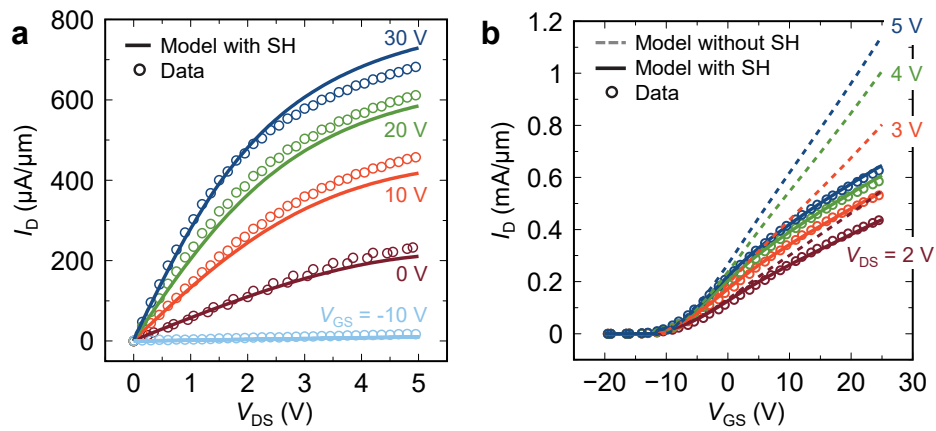

**Fig. 4. High current density and self-heating in AlO$_x$-doped MoS$_2$ FET.** Measured data (symbols) compared to model with self-heating (with SH, lines) and model without self-heating (without SH, dashed). The **(a)** output and **(b)** transfer characteristics correspond to the device in Figure 3, with $L$ = 380 nm. The model includes self-heating with measured thermal conductance.[63] Including self-heating accurately reflects the saturation of current, while the model without self-heating suggests $I_D$ could reach over 1 mA/µm. The simulations also capture the decrease of the saturation voltage ($V_{Dsat}$) with increasing $V_{GS}$, which is only modeled correctly when including self-heating.

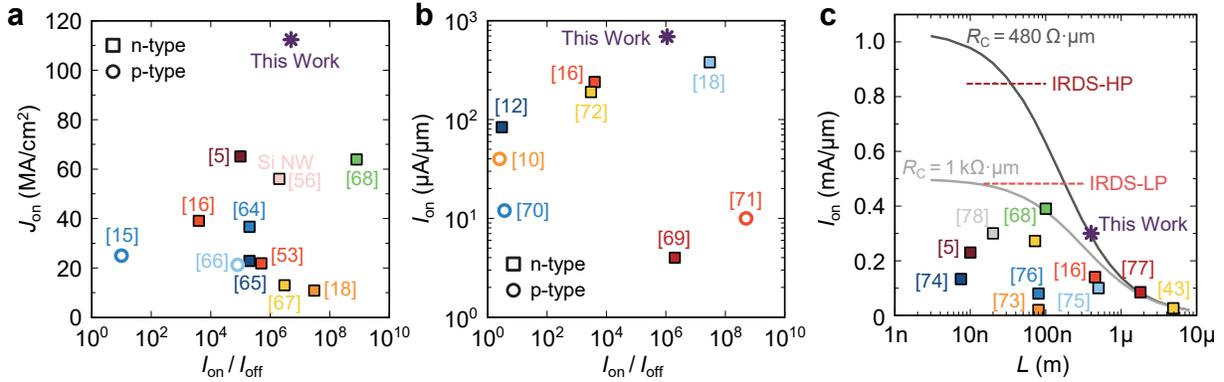

**Fig. 5. Benchmarking various 2D semiconductor transistors. (a)** Maximum current density ($J_{on} = I_{on}/t_{ch}$) vs. $I_{on}/I_{off}$ for our monolayer MoS$_2$, for other high current 2D semiconductor devices [5,15,16,18,53,64-68] of varying layer thickness (0.6 nm ≤ $t_{ch}$ ≤ 14 nm), and a Si nanowire device.[56] **(b)** Maximum $I_{on}$ vs. $I_{on}/I_{off}$ of our AlO$_x$-doped MoS$_2$ compared with other doped 2D-FETs.[10,12,16,18,69-72] While some doping methods yield high $I_{on}$, the $I_{on}/I_{off}$ can be much lower due to charge trapping in the 2D material or its interface. **(c)** Benchmarking of monolayer MoS$_2$ transistors $I_{on}$ vs. channel length ($L$) at $V_{DS}$ = 1 V.[5,16,43,68,73-78] Also shown are IRDS high-performance (HP) and low-power (LP) metrics at $V_{DS}$ = 0.75 V and 10-20 nm gate lengths.[9] The simple model (solid lines) estimates achievable $I_{on}$ with $R_C$ = 480 Ω·μm (this work) or 1 kΩ·μm. The low $R_C$ achieved in this work (or lower) is needed to meet IRDS-HP metrics at reduced channel lengths. The data points are meant to be representative and not exhaustive. A more complete data set is available on our 2D transistor benchmarking website.[58]

Supporting Information for:

# High Current Density in Monolayer MoS$_2$ Doped by AlO$_x$

Connor J. McClellan[1], Eilam Yalon[1,2], Kirby K.H. Smithe[1], Saurabh V. Suryavanshi[1], and Eric Pop[1,3,*]

[1]*Electrical Engineering, Stanford University, Stanford, CA 94305, USA*
[2]*Present address: Electrical Engineering, Technion, Israel Institute of Technology, Haifa, 32000, Israel*
[3]*Materials Science and Engineering, Stanford University, Stanford, CA 94305, USA*

*Corresponding author: epop@stanford.edu

**Section S1. Raman and PL Data on Doping**
We used Raman spectroscopy and photoluminescence (PL) measurements for initial characterization of AlO$_x$-doped MoS$_2$. Figure 1d from the main text displays Raman spectra of MoS$_2$ before and after doping with the AlO$_x$ encapsulation layer. We observe red-shifts in both the E' and A$_1$' peaks after AlO$_x$ deposition. The A$_1$' peak is expected to red-shift with increasing carrier concentration of MoS$_2$,[1] and the observed 0.7 cm$^{-1}$ shift corresponds to induced carrier density $\Delta n \sim 3.2 \times 10^{12}$ cm$^{-2}$, lower than $\Delta n \sim 8.6 \times 10^{12}$ cm$^{-2}$ obtained by electrical characterization. However, we note the electrical measurement is more accurate than the Raman estimate, due to the limited spectrometer resolution.

The E' peak of monolayer (1L) MoS$_2$ is sensitive to strain[2] but its asymmetry seen in main text Fig. 1d is consistent with a doping-induced Fano effect, which has been previously noted in several other semiconductors with high doping.[3,4] Raman measurements of MoS$_2$ capped by 150°C ALD-deposited AlO$_x$ *without* the Al seed layer (Fig. S1) show no shift of the A$_1$' peak and no asymmetry of the E' peak, suggesting that the doping effect is enhanced with the Al seed layer (which subsequently oxidizes) for conformal deposition of AlO$_x$ and doping of the underlying MoS$_2$.

PL measurements in main text Fig. 1c show a decrease and slight red shift of MoS$_2$ PL after AlO$_x$ capping. The shift is consistent with the effects of tensile strain[2] and dielectric screening[5] that decrease the MoS$_2$ optical band gap. In addition, the broadening of the PL peak indicates higher rate of non-radiative recombination, ostensibly due to the presence of charge and defects in the AlO$_x$.

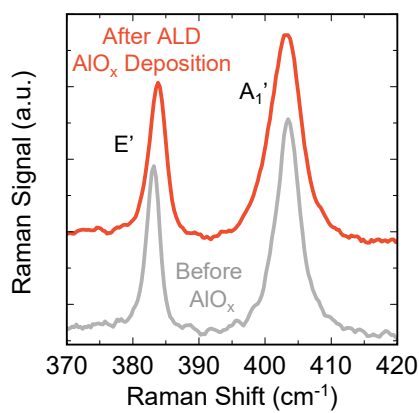

**Fig. S1 Raman Spectra.** Raman spectra of bare MoS$_2$ before AlO$_x$ and after 150°C ALD-AlO$_x$ deposition *without* Al seeding layer. Contrasting with the Raman spectra in main text Fig. 1d, not using the Al seeding layer causes no change in the A$_1$' peak and no Fano asymmetry of the E' peak.

**Section S2. Extraction of Interface Trap Density from Change in Subthreshold Swing**

We estimate the interface trap density ($D_{it}$) with the standard model of subthreshold current in field-effect transistors (FETs), aided by the diagram in Fig. S2. The subthreshold swing is:[6,7]

$$SS \approx (\ln 10)\frac{k_B T}{q}\left(1 + \frac{C_{it} + C_q}{C_{ox}}\right) \tag{E1}$$

where $k_B$ is the Boltzmann constant, $T$ is the temperature, $q$ is the elementary charge, $C_{it}$ is the interface trap capacitance ($C_{it} = q^2 D_{it}$), $C_q$ is the MoS$_2$ quantum capacitance, and $C_{ox}$ is the oxide capacitance ($C_{ox} = \epsilon_{ox}/t_{ox} \approx 115$ nF/cm$^2$ for our 30 nm SiO$_2$ back-gate oxide). We ignore the depletion capacitance ($C_D = \epsilon_{MoS2}/t_{mos2}$) for monolayer MoS$_2$, as it will be much larger than the series $C_q$ and $C_{it}$. As the Al deposition (on top) will not affect $C_{ox}$ (bottom) or $C_q$, we can approximate the change in $D_{it}$ due to AlO$_x$ capping of MoS$_2$ from the change in $SS$, resulting in:

$$\Delta D_{it} \approx \frac{\Delta SS \times C_{ox}}{(\ln 10) k_B T} \tag{E2}$$

This equation stipulates that the $SS$ of MoS$_2$ FETs depends on the interface trap density ($D_{it}$), but is independent of the mobile charge concentration and should not change with doping.

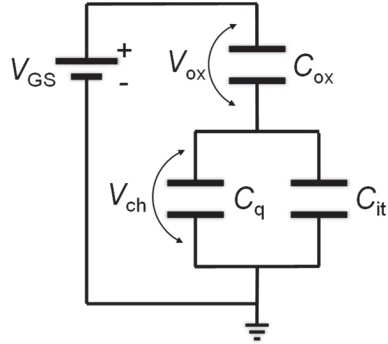

**Fig. S2. MoS$_2$ FET Capacitance Schematic.** Circuit schematic of the monolayer MoS$_2$ transistor including quantum capacitance ($C_q$) and interface trap capacitance ($C_{it}$) effects on the gate control. $V_{ch}$ and $V_{ox}$ are the voltages dropped across the channel and oxide, respectively.

**Section S3. Modeling of Charge Trapping and Charge Doping**

To illustrate the difference between trapping and doping, we use a drift-diffusion model[6] to capture the transfer characteristics of a 2D $n$-type FET with both interface traps and doping. The classical drift current[7] in the linear region describes electron motion from source to drain:

$$I_{drift} = qnv_d \tag{E3}$$

$$n = C_{ox}\left(V_{GS} - V_T - \frac{V_{DS}}{2}\right)/q \tag{E4}$$

$$v_d = \frac{\mu_{eff} F}{\left[1 + \left(\frac{\mu_{eff} F}{v_{sat}}\right)^\gamma\right]^{1/\gamma}} \tag{E5}$$

where $V_T$ is the threshold voltage, $n$ is the average electron density between source and drain, $v_d$ is the average drift velocity in the MoS$_2$ channel, $F$ is the average lateral field $F \approx (V_{DS} - 2I_D R_C)/L$, $v_{sat}$ is the saturation velocity of MoS$_2$ ($\approx 4\times10^6$ cm/s), $\gamma$ is an empirical fitting parameter ($\gamma \approx 5$) and $\mu_{eff}$ is the effective electron mobility.[8]

However, modeling the subthreshold diffusion current is difficult in a monolayer 2D transistor, as the depletion capacitance is large and the quantum capacitance ($C_q$) dominates. Ref. 9 gave equations for $C_q$ with a known Fermi energy ($E_F$), although $E_F$ is not easily estimated in experimental devices. As our goal is modeling the effects of traps on the change in subthreshold current, we estimate $C_q$ from the undoped 2D FET using eq. E1 over a $V_{GS}$ range, and then calculate $E_F$ using Ref. 9. With the extracted $E_F$ for the undoped MoS$_2$ FET, we then incorporate trap levels with delta distributions in computing $C_q$ using eq. 4 of Suryavanshi et al.[6] Thus, the subthreshold diffusion current is:

$$I_{\text{diff}} = \frac{qD_n N_{2D}}{L} \ln\left(\frac{\exp\left[\frac{q(V_{GS}-V_T)}{C_r k_B T}\right]+1}{\exp\left[\frac{q(V_{GS}-V_T-V_{DS})}{C_r k_B T}\right]+1}\right) \quad (E6)$$

where $D_n = (k_B T/q)\mu_{\text{eff}}$ is the electron diffusion coefficient, $N_{2D}$ is the 2D density of states from equation 2 of Suryavanshi et al.[6] and $C_r$ is the normalized capacitance term [$C_r = 1 + (C_q + C_{it})/C_{ox}$]. The total current is then simply $I_D = I_{\text{drift}} + I_{\text{diff}}$ from equations (E3) and (E6) above.

For the simulations shown in the main text Fig. 2 we use the following parameters, also labeled on Fig. S3. For the initial MoS$_2$ device without AlO$_x$ capping we use $V_T$ = 10 V, $\mu_{\text{eff}}$ = 33 cm$^2$V$^{-1}$s$^{-1}$, and a native interface trap density $D_{it}$ = 5×10$^{12}$ cm$^{-2}$eV$^{-1}$ at an energy level $E_{it}$ = -100 meV (i.e. 100 meV below the MoS$_2$ conduction band) to capture the traps already present within the CVD-grown MoS$_2$ or at the SiO$_2$/MoS$_2$ interface. This $D_{it}$ at $E_{it}$ = -100 meV is present in all our simulations, as these traps remain at the SiO$_2$/MoS$_2$ interface.

After deposition of AlO$_x$, we add two trap levels at $E_{it}$ = -200 meV and -50 meV, each with $D_{it}$ = 2.5×10$^{13}$ cm$^{-2}$eV$^{-1}$ without changing other model parameters (i.e. constant $V_T$ and $\mu_{\text{eff}}$). To model the channel after 200°C annealing in N$_2$, we remove the trap levels from $E_{it}$ = -200 meV and -50 meV, but add an additional trap level at $E_{it}$ = 250 meV (above the conduction band) with $D_{it}$ = 7×10$^{12}$ cm$^{-2}$eV$^{-1}$. These parameters are listed on and correspond to the three scenarios labeled in Fig. S3.

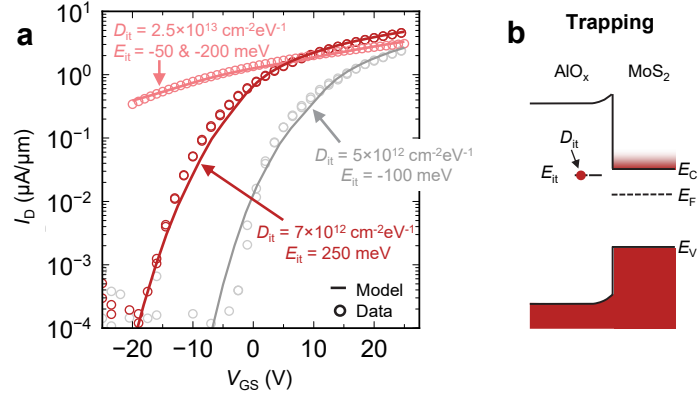

**Fig. S3. Experimental I-V and Trap Modeling. (a)** Measured $I_D$ vs. back-gate $V_{GS}$ (symbols) shown in main text Fig. 2b. Simulations (lines) are shown with parameters used. Colors are consistent with Fig. 2 in the main text: (1) gray is the bare MoS$_2$ device before AlO$_x$ deposition, (2) light red is right after deposition, and (3) dark red is after the anneal step. We note the negative threshold voltage shift after AlO$_x$ deposition (due to doping), and the recovery of the good on/off ratio after the anneal step. **(b)** Band diagram showing $E_{it}$ and $D_{it}$ in the AlO$_x$ on MoS$_2$ in the trapping state, i.e. $E_{it}$ within the MoS$_2$ band gap. The charge trap distribution is incorporated in the model as a delta function $D_{it}\delta(E-E_{it})$, as described by Suryavanshi et al.[6]

### Section S4. Top-Gate Measurements with Doping AlO$_x$ Layer

We also evaluate the AlO$_x$ capping layer as a top-gate dielectric (Fig. S4a). The double-gate transistor has a source-drain contact separation $L$ = 6 μm, top gate length $L_G$ = 5 μm, and channel width $W$ = 3 μm, confirmed by atomic force microscopy (AFM). The Pd top-gate was defined using electron beam lithography and deposited with electron-beam evaporation. After the electron-beam evaporation step, we observed that all devices on the sample displayed trapping-like characteristics (Fig. S4b). The degradation of top-gate control originates from an increase in the density of mid-gap defect states due to oxide damage induced by the high energy X-rays emitted during electron-beam evaporation.[10,11] However, a 40 min 200°C $N_2$ anneal recovers the gate control to the "doping" state (Fig. S4b).

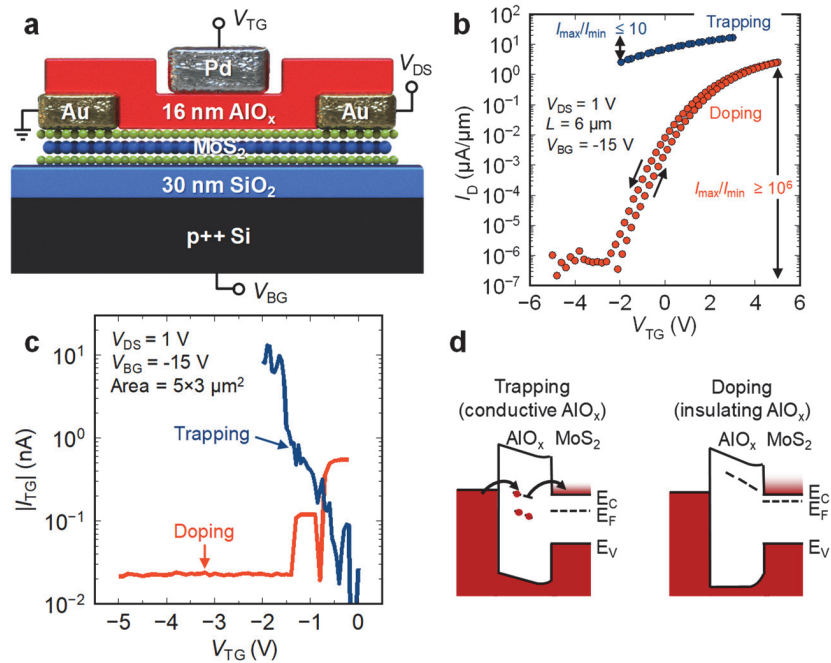

**Fig. S4. Top-Gate Measurements: (a)** Schematic of dual-gated $MoS_2$ FET, using the $AlO_x$ layer as a top-gate dielectric. **(b)** Measured $I_D$ vs. $V_{TG}$ data showing the large $I_{on}/I_{off}$ of the doping state and small $I_{on}/I_{off}$ of the trapping state, similar to data using the back-gate. The arrows mark forward and backward sweeps, indicating relatively low hysteresis. **(c)** Top-gate leakage measurements showing that the $AlO_x$ dielectric with a large trap concentration ("trapping") is more conductive than with low trap concentration ("doping"). **(d)** Energy band diagram showing how mid-gap traps in the $AlO_x$ lead to trap-assisted tunneling and high gate leakage, but higher-energy state traps (in the "doping" state of the oxide) do not.

Fig. S4c displays the measured top-gate leakage current ($I_{TG}$) for the trapping and doping states. Large $I_{TG}$ is measured for the trapping state, limiting the top-gate voltage ($V_{TG}$) sweep from only -2 V to 3 V in Fig. S4b. For the doping state, $I_{TG}$ is reduced to <10 pA, allowing for a $V_{TG}$ sweep from –5 V to 5 V. The large contrast in $I_{TG}$ between trapping and doping offers insight into the state of the $AlO_x$ in these two cases, illustrated with schematic energy band diagrams in Fig. S4d. $AlO_x$ in the trapping state is leaky due to defects that promote electron conduction and trap $MoS_2$ electrons, degrading FET performance and increasing $I_{TG}$ by trap-assisted tunneling. $AlO_x$ in the doping state has higher defect energy levels, above the $MoS_2$ conduction band, reducing trap-assisted tunneling and decreasing $I_{TG}$. The lower $I_{TG}$ indicates that post-anneal $AlO_x$ can be effectively used to dope the underlying 2D semiconductor while also serving as a top-gate dielectric, allowing for process integration of doping and dielectric formation. However, future studies will need to reduce the physical (and equivalent) oxide thickness of the top-gate dielectric, and/or combine it with an additional layer which has a higher dielectric constant (e.g. $HfO_2$). This is needed to reduce the operating gate voltage of $MoS_2$ transistors.

**Section S5. Transfer Length Method Measurements**
We use the transfer length method (TLM)[12] to obtain both sheet and contact resistance of our $MoS_2$ after doping with the $AlO_x$ capping layer. Fig. S5a shows a TLM structure with six channel lengths of 180, 280, 380, 480, 680, and 980 nm (measured by AFM). Fig. S5b plots the total measured resistance ($R_{tot}$) vs. channel length ($L$), showing the expected linear scaling. The sheet resistance ($R_{sh}$) and contact resistance ($R_C$) are extracted from the slope and vertical intercept of the TLM plot as:

$$R_{tot} = R_{sh}L + 2R_C. \quad (E7)$$

With the extracted $R_{sh}$, the effective mobility ($\mu_{eff}$) is obtained as:

$$\mu_{eff} = (qnR_{sh})^{-1} \quad (E8)$$

where the carrier density $n$ is estimated from the gate voltage in eq. E4. We note that due to uncertainties in the threshold voltage $V_T$ (and due to small contributions from $C_q \leq 5\%$ at $n \geq 5 \times 10^{12}$ cm$^{-2}$) the carrier density $n$ and therefore the mobility are more accurately estimated at larger $V_{GS}$. There are also small $V_T$ variations between the different channels within the TLM structure, and thus the TLM extraction is performed at the same gate overdrive ($V_{GS} - V_T$) for each individual channel. Additional details of TLM extraction, uncertainty estimates, and other pitfalls are given in English *et al.*[12]

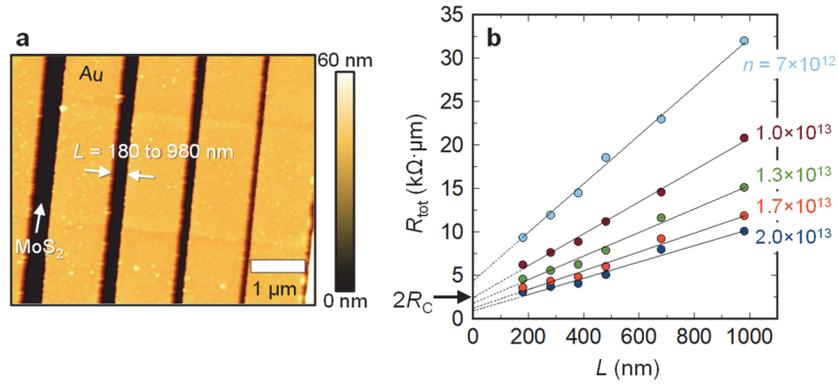

**Fig. S5. TLM Measurements: (a)** AFM of TLM structure on monolayer MoS$_2$, from main text Fig. 1b. **(b)** Measured $R_{tot}$ vs. $L$, used for extracting $R_C$ and $R_{sh}$ at different carrier densities. All lengths were measured by SEM and AFM, confirming channel lengths ~20 nm smaller than target values (i.e. $L$ = 980 nm, 680 nm, 480 nm, etc.). The 380 nm channel had slightly better characteristics than other channel lengths (i.e. lower $V_T$) while 680 nm was slightly worse, causing some of the uncertainty in the $R_C$ extraction.

The effective mobility $\mu_{eff}$ may be underestimated vs. the Hall mobility because the extraction of $n$ may be overestimated due to traps in the MoS$_2$ and/or surrounding dielectrics,[13,14] as discussed in the main text. However, $\mu_{eff}$ is an effective mobility that captures how well the gate controls $R_{sh}$ of the MoS$_2$ (independent of $R_C$), and is also used to calculate the current with eqs. E3-E5. Thus, $\mu_{eff}$ is the correct metric which captures the transconductance and net current flow in these transistors.

From the TLM data, we can also extract the current transfer length ($L_T$), which is the characteristic distance that electrons travel in the semiconductor under the metal contact before flowing up into the metal. This can be simply estimated as $L_T = R_C/R_{sh,c}$ where $R_{sh,c}$ is the sheet resistance of the MoS$_2$ under the contact.[12] For simplicity, we use our average channel $R_{sh} \approx 9$ k$\Omega$/□ but the actual $R_{sh,c}$ could be higher due to (some) metal evaporation damage to the MoS$_2$ under the metal contact. From this, we estimate an upper bound of $L_T$ = 53 nm at $n = 2\times10^{13}$ cm$^{-2}$, indicating the contact length of our devices could be scaled to ~50 nm before contact current crowding effects become non-negligible.

**Section S6. High On-Current and High On/Off**
While achieving high drain current in transistors can decrease circuit delay, transistors must also have a high $I_{on}/I_{off}$ ratio to maintain low leakage current. Fig. S6 plots the measured log-scale $I_D$ vs. $V_{GS}$ of a doped MoS$_2$ FET showing $I_{on}/I_{off} > 10^6$ at both $V_{DS}$ = 2 V and 5 V. These results contrast many previous reports of high current in 2D material transistors, where increasing lateral field (i.e. $V_{DS}$) results in an exponential increase in $I_{off}$, reducing $I_{on}/I_{off}$. The increase in $I_{off}$ is common in small band gap material transistors, such as black phosphorus, where larger lateral field increases band-to-band leakage current. Our devices can maintain a high $I_{on}/I_{off}$ as monolayer MoS2 has a larger band gap ($E_G$ > 2 eV)[15] reducing band-to-band tunneling effects.

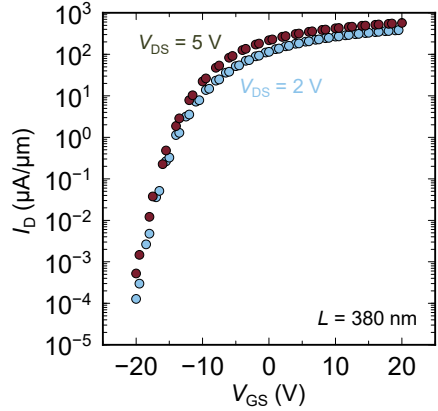

**Fig. S6. High-Current $I_D$-$V_{GS}$:** Measured log $I_D$ vs. $V_{GS}$ of a highly doped MoS2 FET showing $I_{on}/I_{off} > 10^6$ at $V_{DS}$ = 2 V and 5 V. The channel length $L$ = 380 nm and back-gate oxide thickness $t_{ox}$ = 30 nm.

## Section S7. Thermal Modeling of MoS2 FETs

We use a compact thermal model of 2D material FETs to estimate the effects of self-heating on device performance. We first calculate the thermal conductance per unit length ($g$) and thermal resistance ($R_{th}$) from MoS2 to the Si substrate back-side from the equations:[6]

$$g = \frac{R_{Cox}}{W} + \left\{ \frac{\pi \kappa_{ox}}{\ln\left[6\left(\frac{t_{ox}}{W}+1\right)\right]} + \frac{\kappa_{ox}}{t_{ox}} W \right\}^{-1} + \frac{1}{2k_{si}} \left(\frac{L}{W_{eff}}\right)^{\frac{1}{2}} \qquad (E9)$$

$$R_{th} = \frac{1}{gL} \qquad (E10)$$

where $R_{Cox}$ is the thermal boundary resistance between MoS2 and SiO2, $W$ is the width of the MoS2 channel, $\kappa_{ox}$ is the thermal conductivity of SiO2, $t_{ox}$ is the thickness of the SiO2, $\kappa_{si}$ is the thermal conductivity of the highly doped Si substrate, $W_{eff}$ is the effective width of the MoS2 device including thermal spreading[16] into the SiO2 ($W_{eff} \approx W + 2t_{ox}$) and $L$ is the length of the MoS2 channel. With an estimation of $R_{th}$, the increase in temperature can be expressed as:

$$T_{avg} = T_0 + PR_{th}\left\{\frac{1+gL_H R_T x - 2xL_H/L}{1+gL_H R_T x}\right\} \qquad (E11)$$

where $T_0$ is the ambient temperature (~295 K for our measurements unless otherwise stated), $P$ is the input power [corrected for the voltage drop across the contacts, $P = I_D(V_{DS} - 2I_D R_C)$], $R_T$ is the thermal resistance into the 35 nm thick Au contacts,[16] $x = \tanh[L/(2L_H)]$, and $L_H$ is the thermal healing length along the MoS2 and to the metal contacts. We estimate $L_H \approx 110$ nm using the equation:

$$L_H = \sqrt{\kappa_{eff} t_{MoS2} \left(\frac{W}{g} + R_{Cox}\right)}, \qquad (E12)$$

where $\kappa_{eff} = \kappa_{MoS2} + \kappa_{cap}(t_{cap}/t_{MoS2})$ is the effective lateral thermal conductivity[17] accounting for parallel heat flow along the MoS2 and the AlOx capping layer ($t_{cap} \approx 15$ nm and $\kappa_{cap} \approx 3$ Wm$^{-1}$K$^{-1}$).[18] Thus, the thermal model includes both heat sinking to the substrate (most important in longer channels, $L > 3L_H$ ~ 330 nm) and heat sinking to the contacts (more important in the shorter channels, $L < 3L_H$ ~ 330 nm). Additional details about this thermal model can be found in previous work.[6,8,17]

| Parameter | Value | Reference |
|---|---|---|
| $R_{Cox}$ | $7 \times 10^{-8}$ m²KW⁻¹ | 19 |
| $\kappa_{ox}$ | 1.3 Wm⁻¹K⁻¹ | 20 |
| $\kappa_{si}$ | 95 Wm⁻¹K⁻¹ | 19 |
| $\kappa_{MoS2}$ | 34 Wm⁻¹K⁻¹ | 21 |

**Table S1. Thermal Parameters** used for our calculations, all near 300 K. $R_{Cox}$ is thermal boundary resistance, i.e. the inverse of the thermal boundary conductance (TBC). The thermal conductivity of silicon ($\kappa_{si}$) corresponds to our highly doped substrates, as measured in Yalon *et al.*[19]

The effect of temperature on the *I-V* characteristics is included through:

$$\mu_{eff} = \mu_0 \left(\frac{T}{T_0}\right)^\beta \quad (E13)$$

where $\beta$ = -1.24 is extracted from temperature-dependent measurements of MoS₂ mobility[8] and $\mu_0$ is the effective mobility at 295 K (~33 cm²V⁻¹s⁻¹ is this work). For thermal conductivities and boundary resistances, we used the values listed in Table S1. The model results shown in the main text Fig. 4 demonstrate how self-heating can significantly limit the on-state current. Thus, improved heatsinking will reduce mobility degradation and velocity saturation,[8] improving overall device performance.

We also estimate the performance of our devices with a 3× reduction of thermal boundary resistance ($R_{Cox}/3$) in Fig. S7. The maximum $I_{on}$ increases to 930 μA/μm at $n = 2 \times 10^{13}$ cm⁻² and $V_{DS}$ = 5 V, as the max temperature decreases from 700 K to 500 K with the improved heatsinking. There is also a more linear increase in current with $V_{GS}$ as the device approaches the $v_{sat}$-limited regime and the current saturation is less dominated by self-heating. Improvement in device thermal resistance can be achieved by using more thermally conductive dielectrics (e.g. h-BN, AlN)[22] or improved thermal interfaces (lower $R_{Cox}$), because the intrinsic thermal conductivity of the MoS₂ plays only a small role. Although shorter channel transistors will have higher power density, decreasing the channel length ($L < 3L_H$) should reduce the overall self-heating by increasing heat sinking to the contacts.

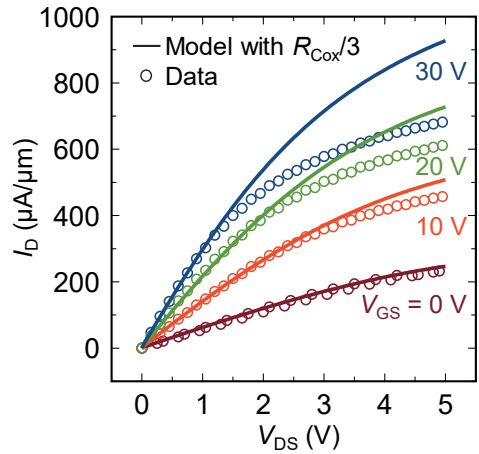

**Fig. S7. Model with Improved Heat Sinking.** Measured $I_D$ vs. $V_{DS}$ data (symbols) of a high-current monolayer MoS₂ FET ($L$ = 380 nm) and simulations (lines) showing that reducing the MoS₂-SiO₂ thermal boundary resistance ($R_{Cox}$) increases the maximum current drive by reducing self-heating. $R_{Cox}$ was reduced from $7 \times 10^{-8}$ m²KW⁻¹ (default parameter in Table S1) to $2.3 \times 10^{-8}$ m²KW⁻¹, corresponding to increased thermal boundary conductance (TBC = $1/R_{Cox}$) from 14.3 MWm⁻²K⁻¹ to 43 MWm⁻²K⁻¹.

**Section S8. Effect of Al Seed layer on Doping**

We also studied the effect of electron-beam physical vapor deposited (EBPVD) Al layers on our CVD-grown monolayer $MoS_2$ FETs. We deposited a series of 1.5 nm layers of Al at a pressure of $10^{-6}$ Torr on $MoS_2$ FETs, exposing the Al to atmosphere for several hours and measuring the electrical characteristics (in a vacuum probe station) between each Al deposition. From previous studies,[23] we know the thin layer of EBPVD Al will completely oxidize upon air exposure, forming a sub-stochiometric $AlO_x$ compound. Thus, this $AlO_x$ layer capping the $MoS_2$ channel is not conductive, and the current is entirely carried by the $MoS_2$.

Fig. S8 shows the *n*-type doping effect of the thin $AlO_x$ layers on the electrical characteristics of the $MoS_2$ FET after the first two 1.5 nm of Al depositions (1.5 nm and 3 nm). However, after further Al depositions (4.5 nm and 6 nm), we see a *decrease* in $MoS_2$ conductivity as the $AlO_x$ doping effect begins to degrade from continued exposure to air. We believe this degradation in conductivity results from carbon contamination on the $AlO_x$ surface and decrease in the fixed charge doping density in the $AlO_x$ layers.[24] This degradation in device performance with Al seeding alone contrasts the stable doping observed with seed layer *followed* by ALD-deposited $AlO_x$, as the higher quality and thicker ALD $AlO_x$ prevents carbon contamination to the $AlO_x$/$MoS_2$ interface. Thus, we conclude that a thin (sub-3 nm) seed layer of Al is necessary but not sufficient to induce the maximum doping effect observed in our $MoS_2$ devices.

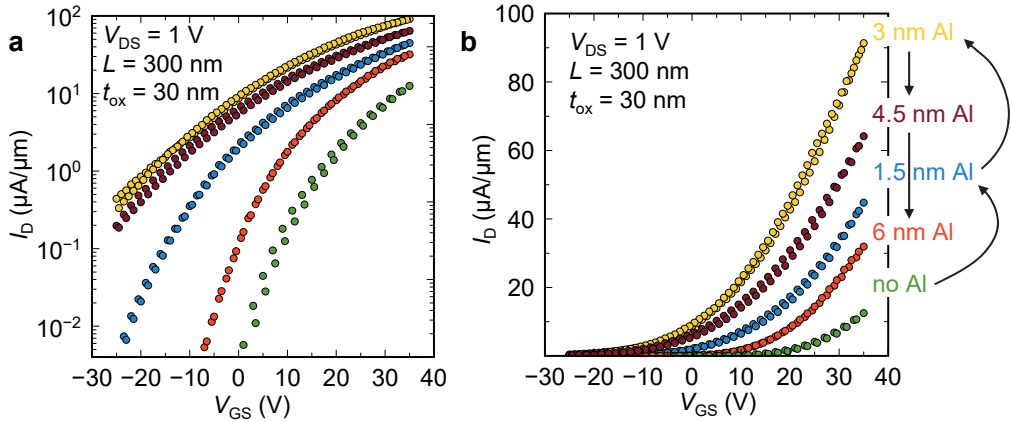

**Fig. S8. Effect of Al Seed Layer.** Measured **(a)** log and **(b)** linear scale $I_D$ vs. $V_{GS}$ of an $MoS_2$ device after a series of Al seed layer depositions (see arrows), showing increase in conductivity from no Al to 1.5 and 3 nm of Al. However, after 4.5 and 6 nm of Al, the conductivity degrades, likely due to surface contamination. The sample is exposed to air after each deposition round, ensuring the $AlO_x$ formed is not conducting. Double sweeps (forward and backward) are shown for each data set, revealing minimal hysteresis.

**Section S9. $H_2$ Annealing to Increase Defects**

While $N_2$ annealing can reduce $MoS_2$/$AlO_x$ interface traps, we found that $H_2$ annealing can increase the interface traps. Fig. S9 demonstrates how the doping concentration can be further increased by a combination of $H_2$ and $N_2$ anneals on an $MoS_2$ device doped with 15 nm ALD-capped $AlO_x$. $H_2$ anneals promote the generation of oxygen vacancies in $AlO_x$ by reacting with oxygen and reducing $AlO_x$.[25] After a 30 minute 150°C Ar/$H_2$ anneal (5% $H_2$), the measured $I_D$ vs. $V_{GS}$ shows trapping, as indicated by the significant increase in *SS*. An $N_2$ anneal increases the trap energy levels, shifting the $V_T$ by -10 V (to more negative values) and increasing the induced carrier concentration by $\Delta n \sim 6\times10^{12}$ cm$^{-2}$ compared to the FET before $H_2$ annealing.

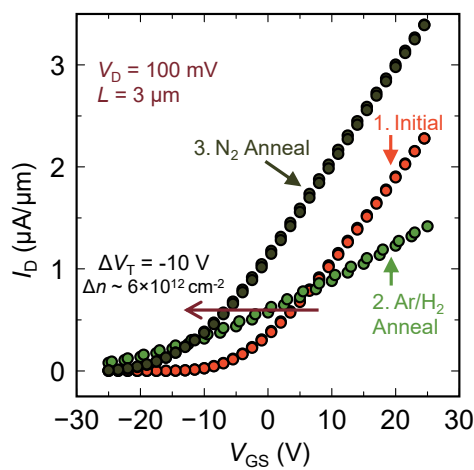

**Fig. S9. $H_2$ Annealing:** Measured $I_D$ vs. $V_{GS}$ data of a $MoS_2$ device doped by ~15 nm ALD-deposited $AlO_x$ capping before treatment, after $Ar/H_2$ annealing, and after $N_2$ annealing. A clear negative threshold voltage $V_T$ shift is observed, indicating the series of $Ar/H_2$ and $N_2$ anneals can increase doping.